\documentstyle[aps,prl,twocolumn,psfig]{revtex}
\draft
\begin{document}
\newcommand {\be}{\begin{equation}}
\newcommand {\ee}{\end{equation}}
\newcommand {\bea}{\begin{eqnarray}}
\newcommand {\eea}{\end{eqnarray}}
\newcommand {\nn}{\nonumber}

\twocolumn[\hsize\textwidth\columnwidth\hsize\csname@twocolumnfalse%
\endcsname

\title{Quantum Antiferromagnetism in Quasicrystals} 

\author{Stefan Wessel$^1$, Anuradha Jagannathan$^2$, and Stephan Haas$^2$}
\address{$^1$Institut f\"ur Theoretische Physik, ETH-H\"onggerberg, 
CH-8093 Z\"urich, Switzerland\\ 
$^2$Department of Physics and Astronomy, University of Southern
California, Los Angeles, CA 90089-0484
}

\date{\today}
\maketitle

\begin{abstract}
The 
antiferromagnetic Heisenberg model is studied on a two-dimensional bipartite
quasiperiodic lattice.
The distribution of local staggered magnetic moments is determined
on finite square approximants with up to 1393 sites,
using the Stochastic Series Expansion Quantum Monte Carlo method.
A non-trivial inhomogeneous ground state
is found.
For a given local coordination number, the values of the magnetic moments are
spread out, reflecting the fact that no two sites in a quasicrystal
 are identical. 
A hierarchical structure in the values of the
moments is observed which arises from the self-similarity of the quasiperiodic
lattice.
Furthermore, the computed  spin structure factor shows antiferromagnetic 
modulations that  can be measured in 
neutron scattering and nuclear magnetic resonance experiments.
 This generic model
is a first step
towards understanding magnetic quasicrystals
such as the recently discovered Zn-Mg-Ho icosahedral structure.

\end{abstract}
\pacs{PACS numbers: }
]

Quantum magnetic phases of low-dimensional antiferromagnetic (AF)
Heisenberg systems
are known to show various degrees of disorder caused by 
zero-point fluctuations. 
For example, isolated spin-1/2 chains have quasi-long-ranged antiferromagnetic
order\cite{cloiseaux}, whereas 
two-leg ladders are short-range-ordered\cite{haldane}, and 
true long-range order is found in the two-dimensional 
square lattice\cite{manousakis}.
While these states are homogeneous due to the translational 
invariance
of the underlying lattice, it is interesting to explore how 
non-periodic environments, such 
as provided in quasicrystal structures, affect the 
magnetic properties of quantum magnets. 
Naturally, it can be expected
that instead of a uniform order parameter, such as the staggered
magnetization in a Heisenberg antiferromagnet,
there will be instead a distribution of local order parameters. The
magnitude of each local moment may, for example, strongly
depend on its environment. In a local viewpoint,
the magnetic moments on
sites with a greater number of neighbors are expected to be suppressed due to
increased local spin fluctuations.
\cite{bulut}
This picture will be modified by the quasicrystalline equivalent of
non-local spin-wave excitations,
leading to a non-trivial distribution of the
order parameters. 
Recent inelastic neutron scattering experiments on the Zn-Mg-Ho
isocahedral quasicrystal \cite{sato} have revealed an antiferromagnetic
superstructure, which fits very well with the antiferromagnetic Heisenberg
model. As opposed to magnetic quasicrystals with itinerant
charge carriers\cite{trambly}, the electrons appear well localized in this material,
and display an antiferromagnetic modulation with a large wave
vector at temperatures below 6 K, similar to the pattern
of the generic quasicrystal structure reported here.

In this work we
use the recently developed
Stochastic Series Expansion Quantum Monte Carlo (SSE-QMC) 
algorithm\cite{sandvik}
to analyze the magnetic ground state properties of 
the nearest-neighbor AF spin-1/2 Heisenberg model,
\bea
H = J \sum_{\langle i,j \rangle} {\bf S}_i \cdot {\bf S}_j,
\eea
on a two-dimensional quasiperiodic tiling. This bipartite 
structure, shown in Fig. 1, is called ``octagonal tiling"
due to its overall 8-fold 
symmetry. Sites in this tiling
 have coordination numbers $z$ ranging from
3 to 8. This generic lattice structure was chosen 
to ensure that the magnetic ground state is unfrustrated due to the bipartite
property, i.e. nearest-neighbor sites belong to two distinct
subtilings.\cite{footnote1}
In the numerical study, approximants with 
$N =41, 239 $, and
$1393$ sites  \cite{duneau} are considered, and 
toroidal boundary conditions are applied.\cite{jagannathan2}  
The temperature is
chosen low enough to obtain the ground state properties of these finite systems.

\vspace{1.0cm}
\begin{figure}[h]
\centerline{\psfig{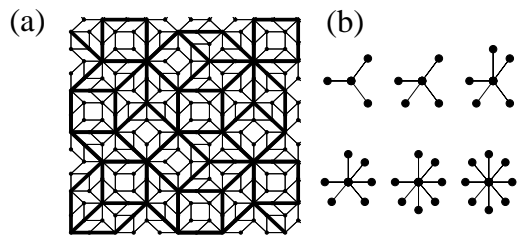}}
\vspace{.5cm}
\caption{Approximants of the octagonal tiling. 
(a) Inflation from 239 (thin lines) to 41 (thick lines) sites.
(b) ``Heisenberg stars" are the local building blocks.
}
\end{figure}

An important property
of quasicrystals, in the absence of invariance under translations,
 is their self-similarity under
inflation transformations. These are site 
decimation operations 
that increase the length scale but globally
preserve the quasiperiodic
structure. An example of inflation is shown in Fig. 1(a) where original 
tiles and inflated tiles are superimposed.
Since periodic approximants are finite pieces of the
octagonal tiling, they are
not invariant under such a transformation, but instead transform into
smaller approximants. In this process, the sites which remain after
an inflation operation can change their connectivity. 
A site of coordination
number z will be transformed into a site of new coordination number $z'$ after
inflation, with $z' \leq z$. We are interested in how this
self-similarity is reflected in the inhomogeneous magnetic 
ground state of the Heisenberg Hamiltonian.

In the SSE-QMC simulations the local value
of the staggered magnetic moment is determined
at each lattice site $i$. It is defined by 
\bea 
m(i)=\sqrt{\frac{3}{N}\sum_{j=1}^{N} (-1)^{i+j} 
\langle S^z_i S^z_j \rangle},
\eea
where the sum extends over all lattice sites j of the approximant.
The inset of Fig. 2
shows a finite-size extrapolation of the spatially averaged 
staggered magnetic moment, $\overline{m} = \sqrt{(3/N^2)\sum 
 (-1)^{i+j}\langle S^z_i S^z_j \rangle}$, where the sum extends over all
pairs of sites $i$ and $j$. 
In the thermodynamic limit, this observable 
approaches $\overline{m} = 0.337 \pm 0.002 $, indicating that the system
has AF long-range order.
Note that the average moment is larger than
that of the square lattice,  $\overline{m}> m_s = 0.3071 \pm 0.0003$
\cite{squarelattice}, even though the average connectivity for the
octagonal tiling is exactly $\bar{z} = 4$, same as for the square lattice. 
This suggests that quantum fluctuations reducing the order
parameter are suppressed 
due to the inhomogeneous connectivity in the octagonal tiling.

\begin{figure}[h]
\centerline{\psfig{figure=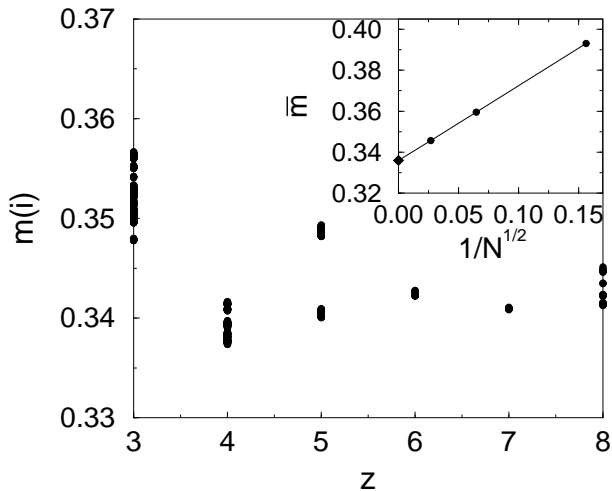,width=8cm,angle=0}}
\vspace{0.0cm}
\caption{
Dependence of the staggered magnetic moments on 
the coordination number $z$.
Inset: finite-size extrapolation of the average staggered magnetic moment. 
}
\end{figure}

The measured dependence of $m(i)$ on the local 
coordination number $z$ 
is plotted in the main part of Fig. 2 for the N=1393 approximant. 
A wide spread of these
moments is observed, particularly for small
values of $z$. 
There is a tendency for sites with more neighbors 
to have smaller staggered moments because their spin is suppressed 
by forming a larger number of local bonds.\cite{bulut} 
However, there are exceptions to this rule, 
indicating the presence of
longer-range correlations. 
A further striking feature observed in Fig. 2 are   
splittings in the distribution function for the staggered moments. This is
most evident at $z = 5$, but it also
occurs on smaller scales at other coordination numbers, e.g. at
$z = 8$.  

Local environments clearly play an important part. To sort out the roles
of the local versus the long-range correlations, we next
focus on the local approximation to the estimator for the staggered
moment, 
\bea
m_{loc}(i)=\sqrt{\frac{3}{z+1}\left(\frac{1}{4}-\sum_{j=1}^{z} \langle S^z_i S^z_j \rangle\right)},
\eea
where the sum is truncated to the sites in the immediate 
vicinity of $i$. This quantity is a local approximation of $m(i)$ defined in
Eq. (2). It reflects the average bond strength $(1/z)\sum_{j=1}^z
{\bf S}_i \cdot {\bf S}_j $ of site $i$ with its $z$ neighbors. 

\begin{figure}[h]
\centerline{\psfig{figure=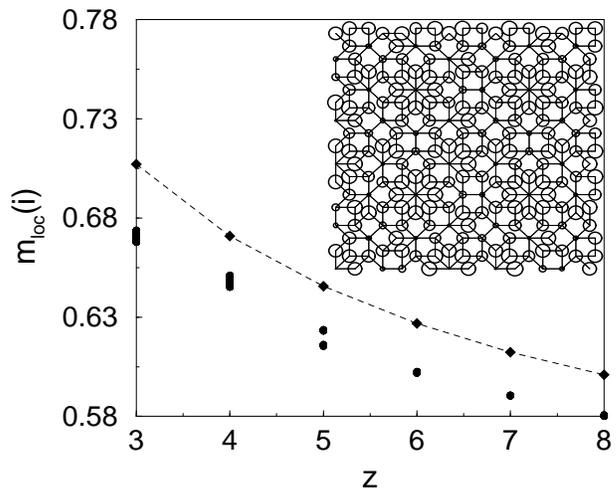,width=8cm,angle=0}}
\vspace{0.0cm}
\caption{
Dependence of the local staggered magnetic moments on 
the coordination number $z$.
The points on the dashed line are exact values obtained for 
Heisenberg stars.
Inset: spatial variation of the
local
approximation to the local staggered moments
where the radii of the circles correspond to the size of local staggered 
moments.
}
\end{figure}

 Fig. 3 shows a clear monotonic decay in $m_{loc}(i)$ 
with increasing number of neighbors $z$. This effect can be understood by
considering the local constituents of the cluster, shown in Fig. 1(b),
which we call Heisenberg stars (HS). These are
described by the nearest-neighbor 
Hamiltonian $H^{HS}(z) = J\sum_{j=1}^{z} {\bf S}_0 \cdot {\bf S}_j$, 
where the index $0$ denotes the central spin of the cluster and
the AF exchange integral is set to  $J = 1$. The ground state energy of
these stars can be calculated exactly, and is given by 
$E^{HS}_0(z) = - (2 + z)/4$. 
The estimator for the local magnetization can be obtained    
directly from Eq. (3),
\bea
m^{HS}_{loc}(z)=\sqrt{\frac{1}{z+1}\left(\frac{3}{4}-E_0\right)} = \sqrt{\frac{5 + z}{4 (z+1)}},
\eea
implying a $z^{-1/2}$-like decay, consistent with the trend of the  QMC data
observed in Fig. 3.
The ground state energy as well as the magnetic moment of the isolated 
Heisenberg clusters are naturally larger than the equivalent quantities 
in the quasicrystal where the presence of additional bonds lowers the 
energy per site and further screens the local magnetization. 

For all coordination numbers, there is some degree of spread in the 
distribution of $m_{loc}(z)$. This phenomenon can be
understood by considering non-local corrections to the magnetic moments of
the isolated Heisenberg stars. 
Next-nearest-neighbor
corrections to the local moments for the
six types of sites ($z= 3,...,8$) can be obtained 
in a self-consistent manner by re-diagonalizing the Heisenberg stars,
using local spin operators that are renormalized according to their specific 
environments, i.e. by replacing $\langle S^z_i \rangle$ by 
$m_{loc}^{HS}(z)$ with the appropriate coordination number 
to determine the renormalized matrix elements.
This procedure naturally leads to larger spreads for 
sites with small $z$, since in the octagonal tiling they have a wider 
variety of local environments than the sites with large $z$,
which is observed in Fig. 3.

Furthermore, one notices a prominent bimodal splitting effect 
at sites with $z = 5$. In the octagonal tiling, these sites
occur in pairs and
have two distinct local environments.\cite{footnote2} 
The first type of z = 5 clusters has two z = 4 neighbors, two z = 3 neighbors,
and one z = 5 neighbor.
The other type has four z = 4 neighbors and one
z = 5 neighbor. Analogous to the discussion in the above paragraph,
the next-nearest-neighbor correction to the staggered magnetic moment
for the two types can be determined 
by diagonalization of the corresponding $6\times6$
matrices in the $S^z_{tot} = (N - 2)/2$ subspace of these two N=6-site 
Heisenberg stars,
using the magnetization values $m_{loc}^{HS}(z)$ of the
Heisenberg stars in Eq. 4.\cite{footnote3} Following this procedure, one finds
two renormalized staggered moments with a splitting $\Delta m_{loc}(5) =
0.007$, which is in very good
agreement with the numerically observed  splitting for the $z=5$ sites in Fig. 3.

Let us now turn to the hierarchical structure of the magnetic ground state 
that can be observed in 
the splittings of the staggered moments of the 
$z = 8$ sites. These sites
have eight-fold symmetry out to their third
neighbor shell. Since their local environments are identical up to
a larger distance compared to other sites,
the spread in values of $m_{loc}(i)$ is relatively
small (Fig. 3). One can however divide these sites into four groups
according to their transformation under inflation, i.e. 
their new local coordination numbers $z'$.
The $m(i)$ values consequently fall into four discrete
groups ($z' = 5,6,7,$ and 8), as shown in Fig. 4. This hierarchical
fine structure is analogous to the discussion of the $z = 5$ sites, but 
occurs on a smaller scale. Interestingly, the most symmetric group
with $z' = 8$ experiences an additional  hyperfine splitting due to a
four-fold subgrouping with $z'' =  5,6,7,$ and 8 under further inflation. 
In the limit of infinite size, this leads to a multifractal distribution of 
$m(i)$ for this class of sites.

\begin{figure}[h]
\centerline{\psfig{figure=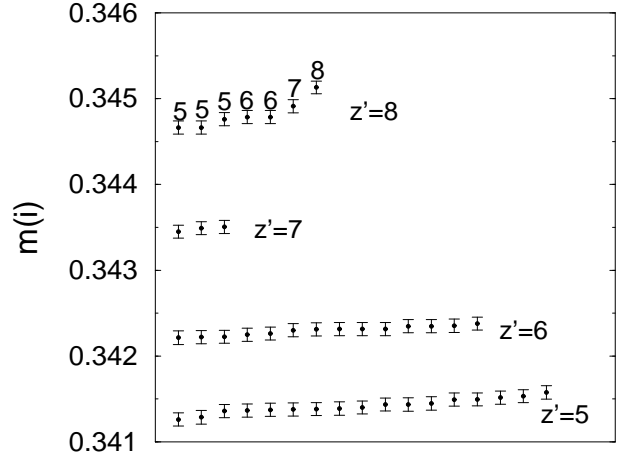,width=8cm,angle=0}}
\vspace{0.5cm}
\caption{Hierarchy of the staggered moment of the $z=8$ sites in the
  1393 site tiling, grouped according to the value of $z'$ under the inflation
  transformation. Numbers on top of the symbols give the value of $z''$ for the
 $z'=8$ sites under a further inflation transformation.
}
\end{figure}

To summarize this discussion,
the different local environments in the quasiperiodic 
structures thus lead, firstly, to a systematic  
decrease of $m_{loc}(i)$ with increasing $z$. 
The differences in next-nearest neighbor shells for a given
$z$ gives rise to a spread of $m_{loc}(i)$ 
and to discrete distributions which are
self-similar on smaller and smaller scales,
as seen in the 8-fold site example. 
 These generic features should be observable 
by high-precision nuclear magnetic resonance measurements. 
Comparing the data in Figs. 2 and 3 one further observes that
long-range correlations tend to suppress the differences
in $m_{loc}(i)$ between the sites,
leading to a non-monotonous dependence of $m(i)$ on $z$.
The spread in values of $m(i)$ is overall smaller
than for $m_{loc}(i)$. However, the
splittings at $z=5$ and $z=8$ persist in $m(i)$, and should thus 
be experimentally observable.

We conclude the discussion of the magnetic ground state on the
octagonal tiling by presenting the magnetic structure factor, 
which is of relevance to inelastic neutron scattering experiments. 
It is obtained by a Fourier transformation of the 
real space correlation function of the largest available approximant,
\bea
S^{zz}({\bf k}) = \frac{1}{N}
\sum_{\bf r,r'} e^{i {\bf k} \cdot ({\bf r} - {\bf r'})}
\langle S^z_{\bf r} S^z_{\bf r'} \rangle,
\eea 
using the QMC data for the octagonal tiling.
In Fig. 5 the diffraction pattern of this quasicrystal 
lattice is compared with the antiferromagnetically induced superstructure.
The eight-fold pattern in Fig. 5(a) reflects the eight-fold symmetry of
the quasicrystal.
The introduction of AF correlations leads
to a splitting of Bragg peaks into the largest available wave vectors
at the Brillouin zone boundary and their symmetry points, as it is seen
in Fig. 5(b). For example, the central Bragg peak splits into 
${\bf Q} \approx (\pm \pi, \pm \pi/2)$ and $(\pm \pi/2, \pm \pi )$, consistent
with the modulation expected for the octagonal tiling.   
This AF superstructure is an extension to the two-dimensional case
of the AF 1D structure discussed in \cite{ron}. 

\begin{figure}[h]
\centerline{\psfig{figure=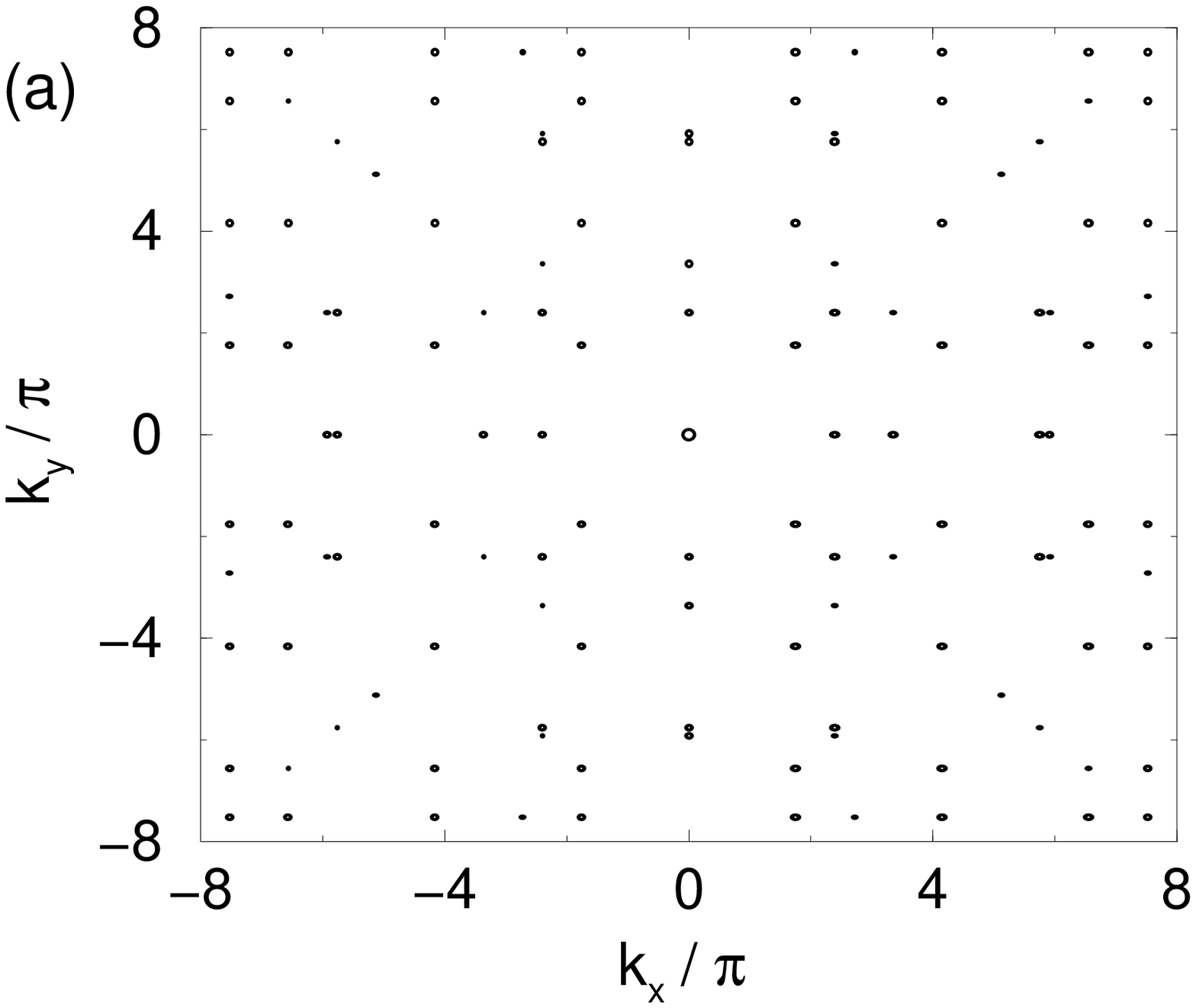,width=6.5cm,angle=0}}
\centerline{\psfig{figure=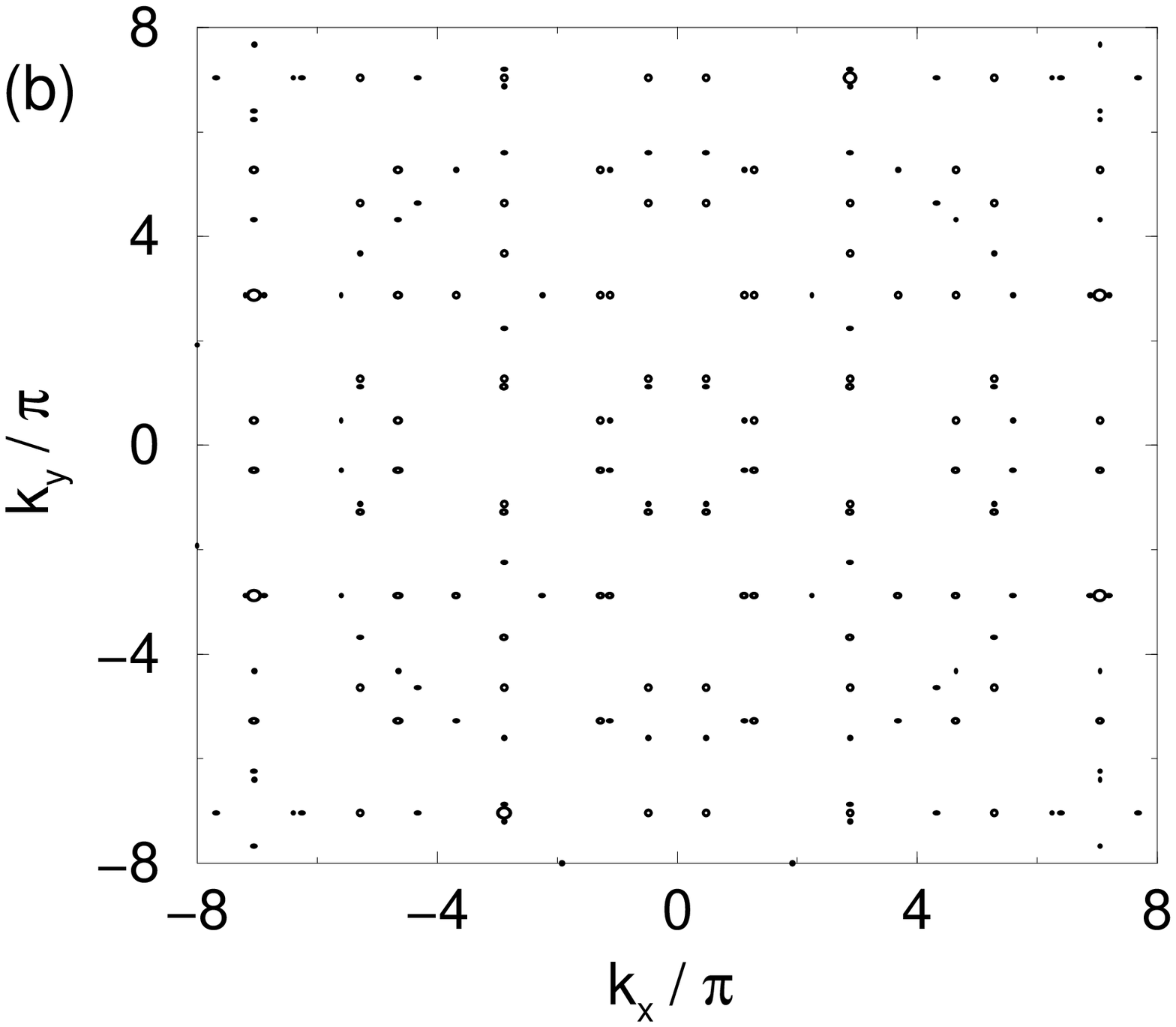,width=6.5cm,angle=0}}
\vspace{.5cm}
\caption{(a) Simple Bragg scattering, and (b) antiferromagnetic 
superstructure in the octagonal tiling, extracted from Quantum 
Monte Carlo data on the 1393 site approximant. 
}
\end{figure}

In conclusion, we have examined the AF spin-1/2 Heisenberg
model on the bipartite octagonal tiling, using the Stochastic Series 
Expansion Quantum Monte Carlo method. 
The main effects we observe 
are (i) an inhomogeneous distribution of staggered moments, (ii)
a reduction of the local order parameter with increasing number of
neighbors, (iii) strong long-range effects,
(iv) evidence for 
hierarchical nature of the magnetic ground state, and (v) a magnetic
superstructure in the spin structure factor. These predictions should 
be tested by high-resolution NMR and neutron scattering measurements. 

We thank
A. L\"{a}uchli, O. Nohadani, B. Normand, R. Roscilde, and M. Sigrist
for useful discussions,
and acknowledge financial support by the Swiss National Fund,
the Petroleum Research Fund, Grant No. ACS-PRF 35972-G6,
and the Department of
Energy,
Grant No. DE-FG03-01ER45908. AJ acknowledges the computational facilities
provided by IDRIS (Orsay). The numerical simulations were performed on the
Asgard Beowulf cluster at the ETH Z\"{u}rich and the hpc cluster at USC.


\begin{references}
\bibitem{footnote1} The present discussion focuses on unfrustrated 
lattices. Magnetic frustrations tend to introduce further disorder in
the ground state, and may even destroy long-range magnetic order.
Moreover, they introduce severe minus-sign problems to
the Monte Carlo algorithm.
\bibitem{footnote2} The members of a pair are inequivalent under
inflation. One site is decimated, while the other remains as a 3-fold
site in the inflated tiling.
\bibitem{footnote3} In this calculation the moments of the 
isolated Heisenberg stars (Eq. 4) are used to determine the
matrix elements. For example, the diagonal entries 
for the cluster with four z = 4 neighbors and one z = 5
neighbor are $H_{ii} = m^{HS}_{loc}(5) [\pm m^{HS}_{loc}(5) \pm 4 
m^{HS}_{loc}(4) ]$.
\end{references}
\end{document}